 \documentclass[twocolumn]{webofc}

\usepackage[varg]{txfonts}
\usepackage{graphicx}
\usepackage{amsmath}
\usepackage{hyperref}
\usepackage{url}
\hypersetup{colorlinks=true,citecolor=blue,urlcolor=blue,linkcolor=blue}
\begin{document}
\title{Enhancing Energy Resolution and Particle Identification via Chromatic Calorimetry: A Concept Validation Study}
%
%

\author{\firstname{Devanshi} \lastname{Arora}\inst{1,2}\fnsep\thanks{\email{devanshi.arora@cern.ch}},
        \firstname{Matteo} \lastname{Salomoni}\inst{1,3},
        \firstname{Yacine} \lastname{Haddad}\inst{4},
        \firstname{Isabel} \lastname{Frank}\inst{1,5},
        \firstname{Loris}  \lastname{Martinazzoli}\inst{1,3},
        \firstname{Marco}  \lastname{Pizzichemi}\inst{1,3},
        \firstname{Michael} \lastname{Doser}\inst{1},
        \firstname{Masaki} \lastname{Owari}\inst{2},
        \firstname{Etiennette} \lastname{Auffray}\inst{1}
        }
\institute{European Organization for Nuclear Research (CERN),Geneva, Switzerland 
\and
           Shizuoka University, Hamamatsu, Japan
\and
           University of Milano-Bicocca. Piazza dell'Ateneo Nuovo, 1, 20126 Milan, Italy
\and
           Northeastern University, Boston, USA
\and          
           Ludwig Maximilian University of Munich, Munich, Germany 
}

\abstract{In particle physics, homogeneous calorimeters are used to measure the energy of particles as they interact with the detector material. Although not as precise as trackers or muon detectors, these calorimeters provide valuable insights into the properties of particles by analyzing their energy deposition patterns. Recent advances in material science, notably in nanomaterial scintillators with tunable emission bandwidths, have led to the proposal of the chromatic calorimetry concept. This proposed concept aims to track electromagnetic or hadronic shower progression within a module, enhancing particle identification and energy resolution by layering scintillators with different emission wavelengths. The idea is to use the emission spectra of the inorganic scintillators to reconstruct the shower progression. Our study validates this proposed concept using inorganic scintillators strategically stacked by decreasing emission wavelength. Using electrons and pions with up to 100 GeV, we achieved analytical discrimination and longitudinal shower measurement. This proof of concept underscores chromatic calorimetry's potential for broader applications.
}
\maketitle
\sloppy
\section{Introduction}
\begin{figure}[ht]
\raggedright
\centering
\includegraphics[width=7cm,clip]{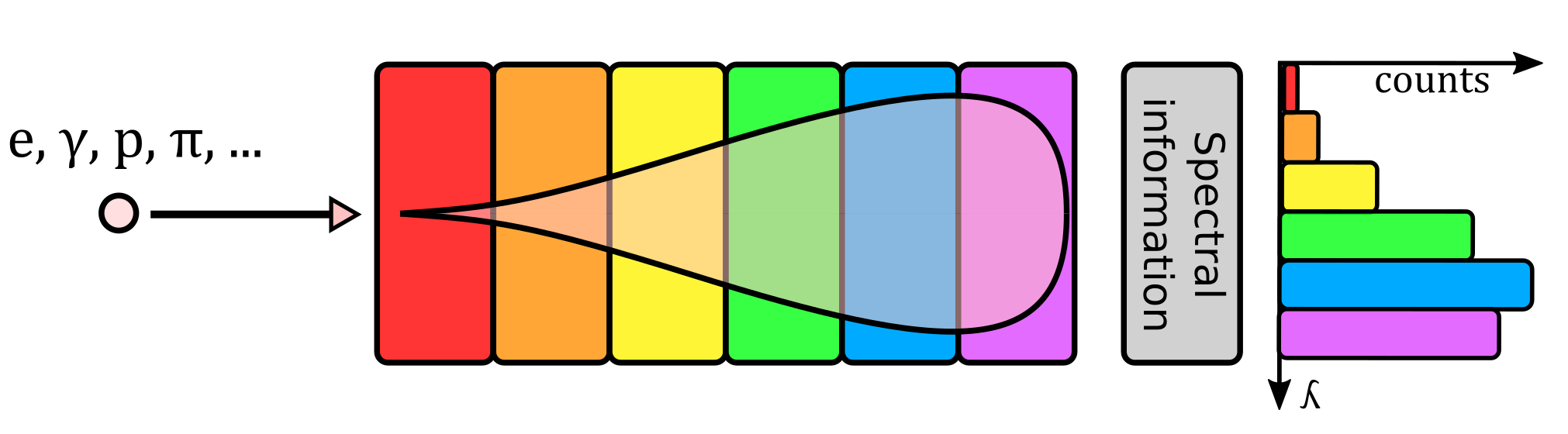}
\caption{Chromatic calorimetry concept: Particle type and energy information can be enhanced by associating each wavelength with a specific longitudinal segment of the calorimeter (modified from \cite{RefB}).}
\label{fig-1}       
\end{figure}

 In particle physics experiments, it is essential to understand the longitudinal evolution of particle showers to identify and characterize incoming particles accurately and to study their interactions with matter. Measuring particle shower profiles in calorimeters plays a vital role in this process. Homogeneous calorimeters, however, lack longitudinal segmentation, making it difficult to obtain detailed information about the shower development along the particle's path. Recent developments in nano-material scintillators have the potential for significant applications in particle physics due to their unique optical and electronic properties \cite{RefJ, RefP}. These nanoscale semiconductors exhibit size-tunable emission spectra, high photo-stability, and brightness, making them highly suitable for advanced detection techniques in calorimetry.
The tunable optical properties of quantum dots  \cite{RefA} allow for better integration into different materials and environments, enhancing their compatibility with other detection system components. These advancements position quantum dots as a pivotal technology in developing next-generation calorimeters for particle physics research. Hence, these advances in the tunability and narrow emission bandwidth ($\sim$20 nm) of quantum wells, carbonized polymer dots, monolayer assemblies, and perovskite nanocrystals enable a new method for measuring the progression of electromagnetic or hadronic showers along a scintillator. This approach offers the potential for longitudinal tomography of the shower profile using a single monolithic device through chromatic calorimetry \cite{RefB}.\\
Chromatic calorimetry represents an innovative approach aimed at enhancing the precision and capabilities of traditional homogeneous calorimeters. This study presents a proof-of-concept device featuring various layers of inorganic scintillating materials with different emission wavelengths. These materials are stacked in the order of decreasing emission wavelengths, with the shortest wavelength positioned at the end of the device. This arrangement enables one-directional transparency, as illustrated in Fig.\ref{fig-1}.
\section{Materials and methods}
\begin{figure}[ht]
\centering
\includegraphics[width=7cm,clip]{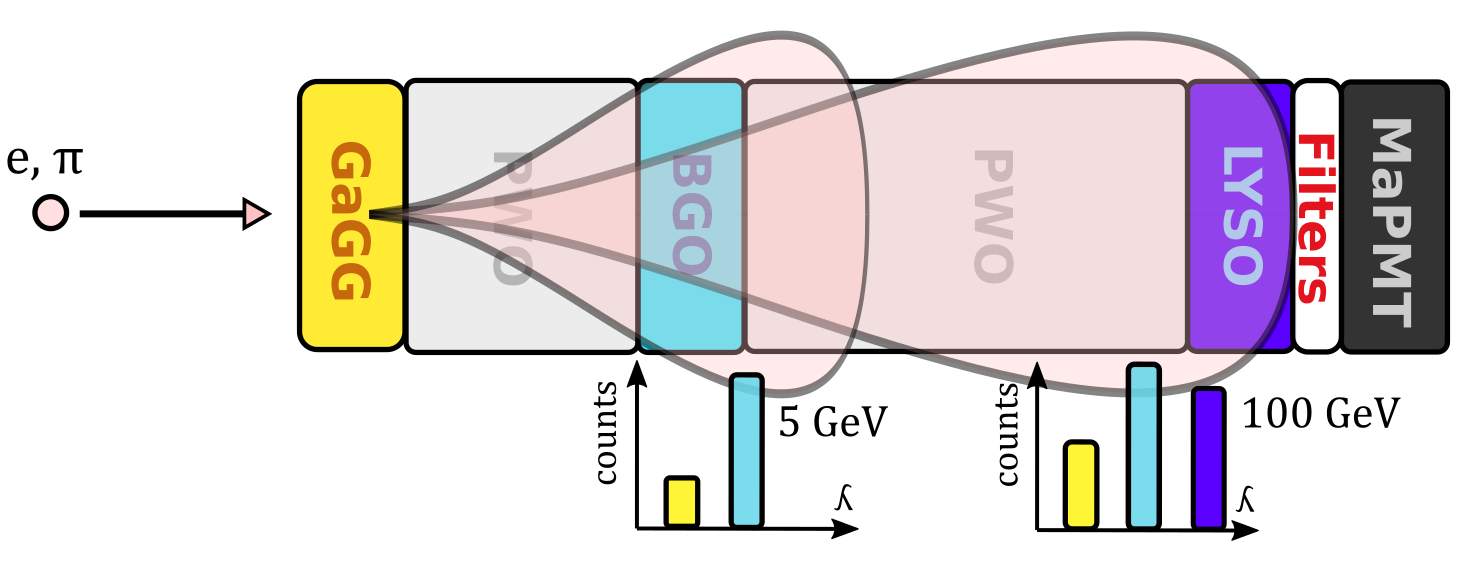}
\caption{Chromatic calorimeter module schematic: The module comprises inorganic scintillators, read out by a MaPMT with four channels. Three different optical filters are used for three channels of the MaPMT: Thorlabs long pass FELH0550 for GAGG, short pass FESH0450 for LYSO, and bandpass FB490-10 for BGO \cite{RefE}. The two histograms display the count number for each scintillator relative to particle penetration depth.}
\label{fig-2}       
\end{figure}

The crystal stack was constructed using the following inorganic scintillators, shown in Fig. \ref{fig-2} (the last dimension is along the longitudinal shower propagation): 1x2x2 cm$^3$ gadolinium aluminum gallium garnet (GAGG, 540 nm peak emission), 2x2x5 cm$^3$ and 2x2x12 cm$^3$ lead tungstate (PWO, 420 nm peak emission), 2x2x3 cm$^3$ bismuth germanate (BGO, 480 nm peak emission), and 2x2x2 cm$^3$ lutetium yttrium oxy orthosilicate (LYSO, 420 nm peak emission). This scheme was designed to obtain scintillation light from 3 crystals: at the beginning of the shower (in GAGG), in the shower maximum (in BGO), and at the end of the shower (in LYSO). PWO crystals, used for their high density and transparency to the emission wavelengths of the other crystals, are integrated within the stack to allow for shower development between GaGG, BGO, and LYSO. The order of the crystal along the shower propagation was chosen to mitigate re-absorption effects, ensuring efficient photon transmission throughout the stack. The experimental scheme was completed with a Hamamatsu Multi-anode Photo-multiplier tube (MaPMT) readout system \cite{RefD}, which selected the emission characteristics of the light emitted at different depths of the shower, thanks to different optical filters (1 per emission of crystal) placed in front of the three out 4 photo-cathode of the PMT, the fourth photo-cathode reading all the signals. \\
We measured energy deposition and photon yield across all output channels using data analysis and Geant4 simulations \cite{RefD1}. These measurements offer valuable insights into the calorimeter's response and the behavior of electromagnetic showers. Our study emphasizes electron-pion discrimination, profiling the shower along its length, and the relationship between energy deposition and the measured amplitude in three different spectral channels. This highlights the need for precise detector design adjustments to improve the effectiveness of chromatic reconstruction.

\subsection{Test Beam Setup}
\begin{figure}[ht]
\centering
\includegraphics[width=7cm,clip]{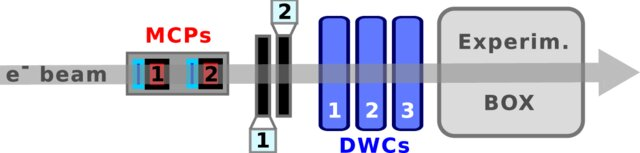}
\caption{Test beam configuration: The electron beam travels from left to right. Two MCPs (Micro-channel Plates) establish the time reference, while two scintillating pads generate the trigger signal. Three Drift Wire Chambers (DWC) supply the tracking information used for the module alignment. The experimental box houses the prototype and the rotating stepper motors (from \cite{RefC}).}
\label{fig-3}       
\end{figure}

The tested chromatic calorimeter module was placed in the experimental box and exposed to a beam of electrons or pions of up to 100 GeV, (see Fig. \ref{fig-3}). The Test beam experiment was performed at the Super Proton Synchrotron (SPS) facility in the zone H2/H6 Northern area of CERN in June 2023 \cite{RefF}.

\section{Results}
\subsection{Study of longitudinal shower profile}
\begin{figure}[ht]
\raggedright
\centering
\includegraphics[width=7cm,clip]{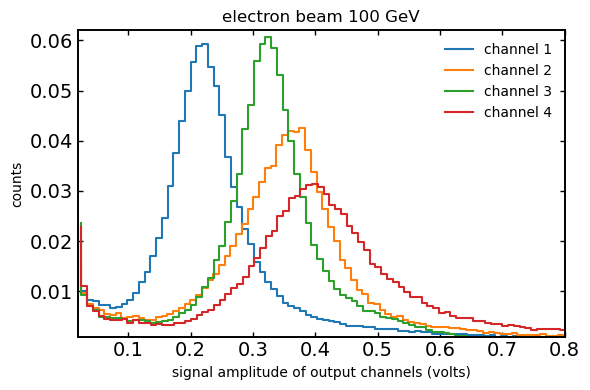}
\caption{Plot illustrating the output signal amplitude measured by the four output channels of MaPMT for 100 GeV electrons, where channels 1, 2, 3, and 4 correspond to the average signal amplitude response from BGO, GAGG, Neutral, and LYSO, respectively.}
\label{fig-5}       
\end{figure}

\begin{figure}[ht]
\raggedright
\centering
\includegraphics[width=7cm,clip]{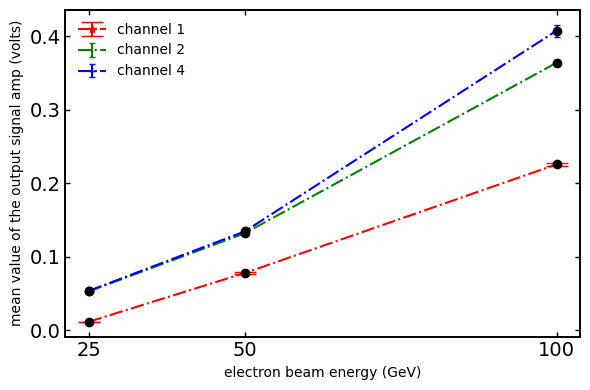}
\caption{The plot illustrates the mean output signal amplitude across three output channels as a function of different electron beam energies. Channels 1, 2, and 4 correspond to average signal amplitude responses from BGO, GAGG, and LYSO respectively. The curve represents results derived from crystal-ball fitting equations.}
\label{fig-6}       
\end{figure}

\begin{figure}[ht]
\raggedright
\centering
\includegraphics[width=7cm,clip]{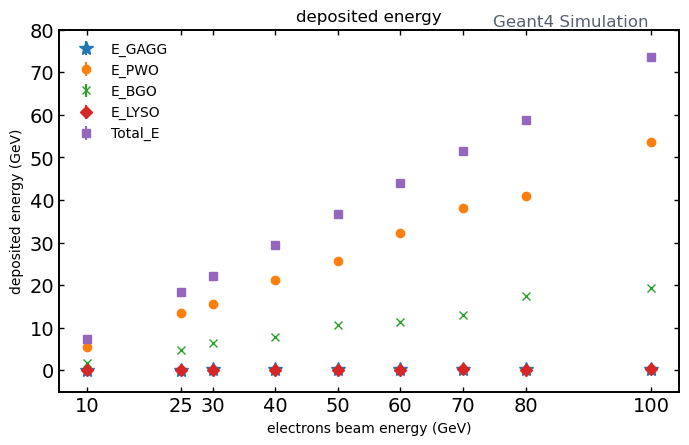}
\caption{Plot showcasing the energy deposited in a calorimeter module composed of GAGG, PWO, BGO, and LYSO materials, as simulated using Geant4 to electron beam energies ranging from 20 to 100 GeV. Here, $E_{GAGG}$ is the energy deposited in GAGG and so on.}
\label{fig-7}       
\end{figure}

From the test beam analysis, output signal amplitudes from various channels of the MaPMT at electron beam energies of 25, 50, and 100 GeV were measured, where channels 1, 2, and 4 tuned to the peak emission from BGO, GAGG, and LYSO, respectively using optical filters. Channel 3 is an additional channel that serves as a "neutral" or "no filter" channel. The measured amplitude spectra for 100 GeV electron beam are shown in Fig. \ref{fig-5}. In this plot, we can see a clear chromatic separation between different channels. As the energy of the electron beam increases, the signals from different channels become more distinct. This separation helps us accurately identify the incoming particles' energy level. Fig. $\ref{fig-6}$ illustrates the mean of the output amplitude for channels 1, 2, and 4 at the same energies (25, 50, and 100 GeV). The mean is determined by fitting the channel amplitude using a Crystal Ball function \cite{RefC1}, which helps to account for the exponential tails. By averaging the signals from these channels, we can see how they respond to changes in energy. The differences in the average amplitudes reinforce the idea of chromatic separation. In Fig. $\ref{fig-7}$, the plot illustrates how much energy is deposited in each layer of the chromatic calorimeter module, as simulated using Geant4. This plot helps us understand how different materials (like GAGG, PWO, BGO, and LYSO) interact with the incoming particles. The amount of energy deposited in each layer affects the signals we get from the MaPMT, linking back to the amplitude spectra we observed in Fig. $\ref{fig-6}$.

\subsection{Electron-Pion Separation}
\begin{figure}[ht]
\raggedright
\centering
\includegraphics[width=7cm,clip]{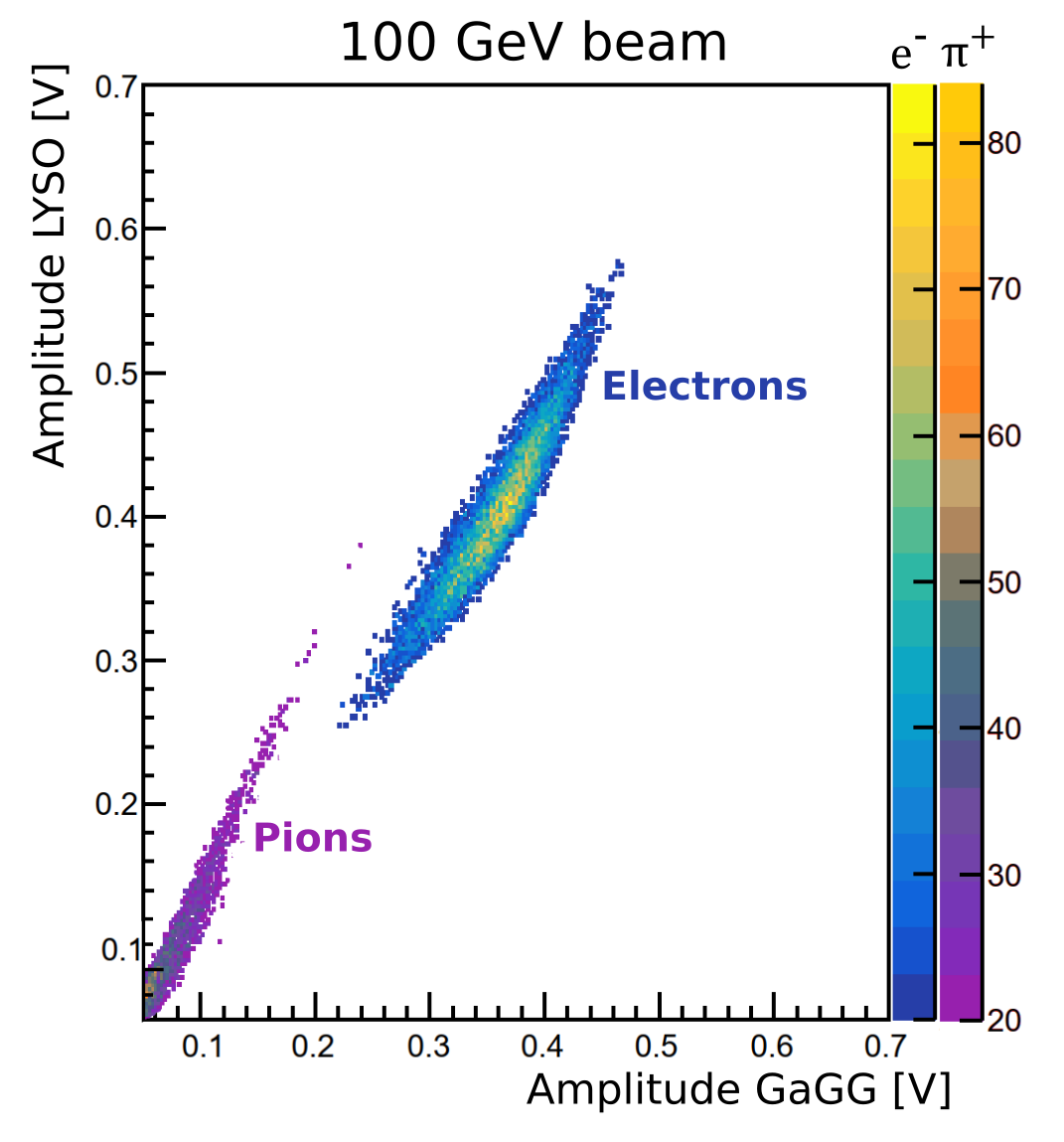}
\caption{Scatter plots display the correlation between signal amplitudes recorded in GAGG and LYSO for electrons (e$^-$) and charged pions ($\pi^+$) at 100 GeV. Each point corresponds to an event, with the x-axis representing the signal amplitude in GAGG and the y-axis representing the signal amplitude in LYSO. Only events with a bin count exceeding 20 are shown.}
\label{fig-4}       
\end{figure}

It was observed that exposing the stack with high-energy electrons or pions at the SPS facility, up to 100 GeV, leads to a change in the amplitude ratio between the crystal responses of GAGG, PWO, BGO, and LYSO. This shift allowed us to distinguish electrons from pions (see Fig. \ref{fig-4}). A similar amplitude scatter plot can be produced for different energies of the same particle.
\\ 
In summary, this prototype, which combines various scintillating crystals with a spectral-sensitive readout scheme (comprising a MaPMT and optical filters), provides an insight into the energy deposited in each sampling layer. A distinct chromatic separation between channels was observed at higher electron beam energies. The mean output amplitudes further confirm this separation, supporting the idea that the electromagnetic shower penetrates deeper into the scintillating material at high energies. However, at lower energies, the separation is less pronounced.

\section{Discussion, conclusion, and outlook}
The test beam analysis demonstrates clear analytical discrimination between electron and pion events. The reconstructed energy signals also reveal valuable separability between hadronic and electromagnetic interactions at higher energies, enhancing our ability to differentiate between these events. Additionally, the signals exhibit a distinct evolution concerning the incoming particle's energy, indicating improved resolution and sensitivity across various energies. These observations highlight the effectiveness of the reconstruction method and its potential for precise event classification.
Chromatic calorimetry aims to capture detailed information on how energy is deposited along the length of a particle shower. The findings demonstrate that a simple arrangement of different types of inorganic scintillators can effectively distinguish between electrons and pions of identical energy levels. This capability comes from the distinct patterns of energy deposition characteristic of each type of particle. However, the experimental setup has revealed some drawbacks. Specifically, the reduced chromatic separation at lower energies can be attributed to two primary factors. First, the geometry of the current prototype is not optimized for clear separation at the shower maximum. Second, the presence of PWO crystals at the front and rear of the module contaminates the spectra from the GAGG and BGO layers. This contamination reduces the effectiveness of chromatic separation and highlights the need for a more refined design.
To address these drawbacks, an optimal setup should employ materials with well-defined emission and absorption spectra, ensuring that each layer contributes uniquely to the detection process. This improved setup is currently under study and will be tested during the 2024 test beam at the SPS facility. Results from a follow-up beam test conducted prior to this submission will be published in a separate paper.\\
The goal of chromatic calorimetry, as described in \cite{RefB}, is to leverage nanomaterials like perovskite and CdSe quantum dots embedded in scintillators with a calorimeter module. With tunable and narrow optical emissions, these materials offer precise control over photon emission, addressing the abovementioned limitations. The approach aims to achieve a more refined longitudinal tomography of electromagnetic showers. 
\section{Author Contributions}
All authors contributed to the paper by providing insight into different sections and correcting the complete document. D.A., E.A., and M.S. planned the design. D.A. performed the data analysis under technical support from Y.H. and M.S. D.A., I.F., L.M., and M.S. conducted the test beam experiment. E.A., M.D., Y.H., M.O., and M.S. guided and supervised the project.
\section{Conflict of Interest}
\raggedright
The authors declare that the research was carried out without any commercial or financial ties that might be seen as a conflict of interest.\\
\section{Acknowledgement}
\raggedright
This work was carried out in the frame of the CERN Quantum Technology Initiative (QTI) and the CERN Crystal Clear Collaboration (CCC). It is also a part of the ECFA-DRD5 collaboration.\\

%
%
%

\end{document}